\documentclass{emulateapj}
\usepackage{graphicx,color,natbib}
\usepackage{amsfonts}
\usepackage[figuresright]{rotating}
\usepackage{ifthen}

\shorttitle{Polarised White Light and Faraday Rotation}
\shortauthors{Xiong et al.}

\begin{document}

\title{Using Coordinated Observations in Polarised White Light and Faraday Rotation to Probe
the Spatial Position and Magnetic Field of an Interplanetary
Sheath}

\author{Ming Xiong \altaffilmark{1, 2}}
\author{Jackie A. Davies \altaffilmark{3}}
\author{Xueshang Feng \altaffilmark{1}}
\author{Mathew J. Owens \altaffilmark{4}}
\author{Richard A. Harrison \altaffilmark{3}}
\author{Chris J. Davis \altaffilmark{4}}
\author{Ying D. Liu \altaffilmark{1}}

\altaffiltext{1}{State Key Laboratory of Space Weather, Center for
Space Science and Applied Research, Chinese Academy of Sciences,
Beijing, China} \altaffiltext{2}{Science and Technology on
Aerospace Flight Dynamics Laboratory, Beijing Aerospace Control
Center, Beijing, China} \altaffiltext{3}{Rutherford-Appleton
Laboratory (RAL) Space,
Harwell Oxford, UK} %
\altaffiltext{4}{Reading University, Reading, UK}

\email{mxiong@spacweather.ac.cn}  

\begin{abstract}
Coronal mass ejections (CMEs) can be continuously tracked through
a large portion of the inner heliosphere by direct imaging in
visible and radio wavebands. White-light (WL) signatures of solar
wind transients, such as CMEs, result from Thomson scattering of
sunlight by free electrons, and therefore depend on both the
viewing geometry and the electron density. The Faraday rotation
(FR) of radio waves from extragalactic pulsars and quasars, which
arises due to the presence of such solar wind features, depends on
the line-of-sight magnetic field component $B_\parallel$, and the
electron density. To understand coordinated WL and FR observations
of CMEs, we perform forward magnetohydrodynamic modelling of an
Earth-directed shock and synthesise the signatures that would be
remotely sensed at a number of widely distributed vantage points
in the inner heliosphere. Removal of the background solar wind
contribution reveals the shock-associated enhancements in WL and
FR. While the efficiency of Thomson scattering depends on
scattering angle, WL radiance $I$ decreases with heliocentric
distance $r$ roughly according to the expression $I \propto
r^{-3}$. The sheath region downstream of the Earth-directed shock
is well viewed from the L4 and L5 Lagrangian points, demonstrating
the benefits of these points in terms of space weather
forecasting. The spatial position of the main scattering site
$\mathbf{r}_{\rm sheath}$ and the mass of plasma at that position
$M_{\rm sheath}$ can be inferred from the polarisation of the
shock-associated enhancement in WL radiance. From the FR
measurements, the local $B_{\parallel {\rm sheath}}$ at
$\mathbf{r}_{\rm sheath}$ can then be estimated. Simultaneous
observations in polarised WL and FR can not only be used to detect
CMEs, but also to diagnose their plasma and magnetic field
properties.
\end{abstract}

\section{Introduction} \label{Sec:intro}
\subsection{The Inner Heliosphere}
The inner heliosphere is permeated with the magnetised solar wind
from the Sun. At solar minimum, the solar wind is inherently
bimodal \citep{McComas2000}, with slow flow tending to emanate
from near the ecliptic and fast flow tending to emanate at higher
latitudes. Several large-scale structures, which pervade
interplanetary space, are associated with the ``ambient" solar
wind: (1) a spiralling interplanetary magnetic field (the Parker
spiral) that forms as a result of solar rotation
\citep{Parker1958}, (2) corotating interacting regions (CIRs) that
are formed at the interface between a preceding slow solar wind
stream and a following fast solar wind stream \citep{Gosling1999},
and (3) the heliospheric current sheet typically embedded in the
heliospheric plasma sheet \citep{Winterhalter1994,Crooker2004}.

The background solar wind flow is frequently disturbed by coronal
mass ejections (CMEs), large-scale expulsions of plasma and
magnetic field from the solar atmosphere. CMEs typically expand
during their propagation, because the total solar wind pressure
decreases with heliocentric distance
\citep{Demoulin2009,Gulisano2010}. The expansion speed of a CME
depends on its spatial size, translation speed, and heliocentric
distance, as well as the pre-existing solar wind conditions
\citep{Nakwacki2011,Gulisano2012}. A number of popular models
describe the motion of a CME as governed by two forces: a
propelling Lorentz force \citep{Chen1989,Chen1996,Chen2006} and an
aerodynamic drag force \citep{Cargill1996,Vrsnak2002,Cargill2004}.
According to these models, the drag force gradually becomes
dominant in interplanetary space, and the CME speed finally
adjusts to the ambient solar wind speed. The equalisation of the
CME and solar wind speed occurs at very different heliospheric
distances, from below 30 solar radii to beyond 1 AU, depending on
the characteristics of the CME and the solar wind
\citep{Temmer2011}. A CME can undergo significant, nonlinear, and
irreversible evolution during its propagation, as it interacts
with the ambient solar wind and other CMEs
\citep[e.g.,][]{Burlaga2002,Demoulin2010}. Coronagraph
observations show that CME morphology is distorted rapidly and
significantly in a structured solar wind
\citep[e.g.,][]{Savani2010,Savani2012,Feng2012a}. Such a
distortion occurs over a relatively short heliocentric distance.
Interaction between multiple CMEs has been revealed by in-situ
observations
\citep[e.g.,][]{Burlaga1987,Wang2003a,Steed2011,Mostl2012}, radio
burst observations \citep[e.g.,][]{Gopalswamy2001,Oliveros2012},
white-light (WL) imaging
\citep[e.g.,][]{Harrison2012,Liu2012,Lugaz2012,Temmer2012,Shen2012a,Bemporad2012},
and numerical magnetohydrodynamic (MHD) simulation
\citep[e.g.,][]{Lugaz2005,Xiong2007,Xiong2009,Shen2012b}.

CMEs cause phenomena at Earth, such as geomagnetic storms and
solar energetic particles, that can result in major space weather
effects \citep{Gopalswamy2006,Webb2012}. Traditionally, a CME has
been defined in terms of a three-part structure, involving a
bright sheath, a dark cavity, and a bright filament. It is now
accepted that the cavity component is an escaping magnetic flux
rope that drives the CME
\citep[e.g.,][]{Rouillard2009b,DeForest2011}. A high-speed flux
rope can drive a fast shock ahead of itself that is much wider in
angular extent than the flux rope itself. The region between the
shock front and the flux rope is defined as a sheath. Within the
sheath, (1) magnetic field lines are draped and compressed, and
(2) a plasma flow is deviated, compressed, and turbulent
\citep[e.g.,][]{Gosling1987,Owens2005,Liu2008b}. Precursor
southward magnetic fields ahead of CMEs are generally compressed,
making them particularly geoeffective
\citep{Tsurutani1992,Gonzalez1999}. The magnetic fields in the
sheath and in the flux rope can be equally important in driving
major geomagnetic storms
\citep{Tsurutani1988,Tsurutani1992,Szajko2013}. In so-called
two-dip storms, it is often the case that the first dip in the
$Dst$ index is produced by the upstream sheath and the second is
produced by the driving flux rope
\citep{Echer2004,Zhang2008,Mostl2012}.

\subsection{Heliospheric White Light Observations}
Heliospheric imagers (HIs) detect WL that has been
Thomson-scattered from free electrons. For resolved objects, such
as CMEs, the power detected by an individual pixel depends
linearly on the solid angle subtended by that pixel ($\delta
\omega$) and the area subtended by the corresponding aperture
($\delta A$), and is proportional to the radiance (measured in W
m$^{-2}$ SR$^{-1}$). The light from unresolved objects, such as
stars, which are much narrower in angular extent, tends to fall
within individual pixels. For a resolved heliospheric electron
density feature, such as a CME, a single pixel provides a measure
of its radiance (surface brightness), while summing contributions
from all pixels over the entire extent of the feature provides a
measure of its intensity (total brightness). The intensity is an
integral of the radiance over the apparent feature size.
Therefore, the feature's intensity determines its detectability of
an object, be it resolved or unresolved \citep{Howard2012}.

The background zodiacal and stellar signals detected by
heliospheric imagers are much more intense than the signal due to
Thomson-scattering from plasma features such as CMEs
\citep{Leinert1989}. Fortunately, using an image-differencing
technique, the much more stable background radiance can be
removed, such that the more transient Thomson-scattering signal
can be extracted. From such processed Thomson-scattering images,
the sunlight-irradiated CMEs can be easily identified and tracked.
According to theory, the heliospheric Thomson-scattering radiance
is governed by the Thomson-scattering geometry factors and
electron number density
\citep{Vourlidas2006,Howard-T-A2009a,Howard2012,Xiong2013}. The
CME detectability in WL is actually more limited by perspective
and field-of-view (FOV) effects than by location relative to the
Thomson-scattering sphere \citep{Howard2012}.

Heliospheric imaging from two vantage points, both off the
Sun-Earth line, was made possible by the Heliospheric Imagers
(HIs) onboard the {\it Solar-TErrestrial RElations Observatory}
({\it STEREO}) \citep{Eyles2009}. With the {\it STEREO}/SECCHI
package, a CME can be imaged from its nascent stage in the inner
corona all the way out to 1 AU and beyond
\citep[e.g.,][]{Harrison2008,Davies2009,Davis2009,Liu2010b,DeForest2011,Liu2013}.
In particular, images from {\it STEREO}/HI-2 have revealed
detailed spatial structures within interplanetary CMEs, including
leading-edge pileup, interior cavities, filamentary structure, and
rear cusps \citep{DeForest2011}. Comparison with in-situ
observations has revealed that the leading-edge pileup of solar
wind material, which is evident as a bright arc in WL imaging,
corresponds to the sheath region. However, the interpretation of
the leading edge of the radiance pattern, especially at larger
elongations, is fraught with ambiguity
\citep[e.g.,][]{Howard-T-A2009a,Xiong2013}. Elongation
$\varepsilon$ is defined as the angle between the Sun-observer
line and a line-of-sight (LOS). Because a CME occupies a
significant three-dimensional (3D) volume, different parts of the
CME will contribute to the radiance pattern imaged by observers
situated at different heliocentric longitudes \citep{Xiong2013}.
Even for an observer at a fixed longitude, a different part of the
CME will contribute to the imaged radiance at any given time
\citep{Xiong2013}. Various techniques have been developed that
enable the spatial locations and propagation directions of CMEs to
be inferred, based on the fitting of their moving radiance
patterns
\citep[e.g.,][]{Sheeley2008,Rouillard2008,Thernisien2009,Liu2010b,Lugaz2010,Mostl2011,Davies2012}.
However, the determination of interplanetary CME kinematics,
propagation direction in particular, are somewhat ambiguous
\citep{Howard-T-A2009a,Davis2010,Davies2012,Howard2012,Xiong2013,Lugaz2013}.

\subsection{Faraday Rotation Measurement}
Faraday rotation (FR) is the rotation of the plane of polarisation
of an incident electromagnetic wave as it passes through a
magnetised ionic medium. The FR observations of linearly polarised
radio sources can be used to estimate magnetic field in the corona
and interplanetary space
\citep[e.g.,][]{Levy1969,Bird1980,Sakurai1994,Liu2007,Jensen2007,
Jensen2008,You2012,Jensen2013}. The FR measurement of a radio
signal corresponds to the path integral of the product of electron
density $n$ and the projection of the magnetic field along the
LOS, $B_\parallel$. The first FR experiment was conducted in 1968
by \citet{Levy1969}, when solar plasma occulted the radio
down-link from the {\it Pioneer} 6 spacecraft. As well as man-made
radio sources, FR experiments can also exploit natural radio
sources such as pulsars and quasars. The first FR experiments of
this type were conducted by \citet{Bird1980} during the solar
occultation of a pulsar. In terms of their locations on a sky map,
many pulsars and quasars lie in the vicinity of the Sun.
Therefore, simultaneous FR measurements along multiple beams can
be used to map the inner heliosphere with a reasonable spatial
resolution.

Additional observations, for example in WL, would generally be
necessary to confirm whether an FR transient was indeed caused by
a CME. For instance, the first FR event, reported by
\citet{Levy1969}, could not be attributed unambiguously to the
presence of any particular solar wind structure. The FR
signatures, observed by \citet{Levy1969}, exhibited a W-shaped
profile over a time period of $2-3$ hours, with rotation angles of
up to 40$^\circ$ relative to the quiescent baseline.
\citet{Woo1997} interpreted the FR signature as the result of a
coronal streamer stalk of angular size $1^\circ - 2^\circ$,
whereas \citet{Patzold1998} argued that it was caused by the
passage of a series of CMEs. However, by comparing observations
from the {\it Solwind} coronagraph and measurements of {\it
Helios} down-link radio signals, \citet{Bird1985} were able to
identify the signatures of five CMEs simultaneously in WL and FR.
Moreover, the electron density derived from WL imaging can be used
to enable magnetic field magnitude to be inferred from FR
measurements.

The heliospheric magnetic field can be remotely probed in FR,
using low-frequency radio interferometers such as the Murchison
Widefield Array (MWA) \citep{Lonsdale2009}, the LOw Frequency
ARray (LOFAR) \citep{deVos2009}, and the Very Large Array (VLA)
\citep{Thompson1980}. Disturbance of the background solar wind by
CMEs will cause the observed FR signatures to become variable
\citep[e.g.,][]{Levy1969,Bird1985,Jensen2008}. A change in either
the electron density ($\delta n$) or the LOS magnetic field
component ($\delta B_\parallel$), or indeed both, will contribute
to the rotation in the plane of polarisation of the radio signal
by $\delta \, \Omega_{\rm RM}$. Interplanetary magnetic clouds
(MCs) in particular, which have a magnetic flux rope configuration
\citep{Burlaga1981,Klein1982,Lepping1990}, can be identified from
WL images \citep{Rouillard2009b,DeForest2011} and are expected to
be easily identifiable in FR measurements. Moreover, $\delta n$
and $|\delta \mathbf{B}|$ are often enhanced simultaneously within
the sheath ahead of a fast MC. The FR due to a MC-driven sheath
can be comparable to that due to the MC itself \citep{Jensen2010}.
It is expected that the orientation and helicity of a MC will be
able to be determined unambiguously from multi-beam FR
measurements \citep{Liu2007,Jensen2007,Jensen2010}. In contrast,
the in-situ detection of magnetic flux ropes can be significantly
hindered by the location of the observing spacecraft
\citep[e.g.,][]{Hu2002,Mostl2012,Demoulin2013}. FR imaging can be
used to provide the magnetic orientation of a fast MC, and indeed
its preceding sheath, prior to its arrival at Earth, which is
crucial for predicting potential space weather effects at Earth.

\subsection{Forward Magnetohydrodynamic Modelling}
Forward modelling of WL and FR signatures is proving extremely
useful for inferring the in-situ properties of interplanetary CMEs
from remote-sensing data. Sophisticated numerical MHD models of
the inner heliosphere
\citep[e.g.,][]{Groth2000,Lugaz2005,Hayashi2005,Xiong2006a,Li2006,Wu2006,Li2008,Odstrcil2009,Shen2012b}
can serve as a digital laboratory, to enable the synthesis of a
variety of observable remote-sensing signatures. In this paper, we
perform a numerical MHD simulation of an interplanetary shock in
the ecliptic, from which we synthesise the signatures of that
feature that would be remotely sensed at visible and radio
wavelengths. Details of the MHD model, and the formulae required
to synthesise the remote-sensing observations, are given in
Section \ref{Sec:method}. The resultant synthesised remote-sensing
signatures of the sheath, which would be observed from vantage
points at 0.5 and 1 AU, are described and compared in Section
\ref{Sec:HalfOneAU}. In Section \ref{Sec:Pattern}, we discuss the
radiance patterns that are observed in the synthesised WL and FR
sky maps. In Section \ref{Sec:Depend}, we explore the role that
the vantage point of the observer plays in the ``observability" of
such WL and FR features. CME detection in the presence of
background noise, and the heliospheric imaging of more complex
interplanetary phenomena, are discussed in Section
\ref{Sec:discussion}. The potentially important role that forward
modelling can play in our understanding of coordinated WL and FR
observations is summarised in Section \ref{Sec:conclusion}.

%------------------------------------------------------------------------------------------------------------------
\section{Method} \label{Sec:method}
Forward MHD modelling can self-consistently establish the links
between interplanetary dynamics and the resultant observable
signatures. A complete flow chart of forward modelling is
illustrated in Figure 8 from \citet{Xiong2011}. The travelling
fast shock studied by \citet{Xiong2013} is revisited here. Our
methodology consists of three general steps: (1) forward modelling
of the shock using the numerical Inner-Heliosphere MHD (IH-MHD)
model \citep{Xiong2006a,Xiong2013}, (2) calculation of its
Thomson-scattered WL signature, in Section \ref{Sec:method-wL},
and (3) calculation of its FR signature, in Section
\ref{Sec:method-FR}. Characterisation of the IH-MHD model, the
background solar wind conditions, and the initial shock injection
is summarised respectively in Tables 1, 2, and 3 of
\citet{Xiong2013}. The simulated electron density $n$ and magnetic
field $\mathbf{B}$ are used to generate synthetic WL and FR
images, which enable us to explore the WL and FR signatures of an
interplanetary sheath.

\begin{figure}
  \includegraphics{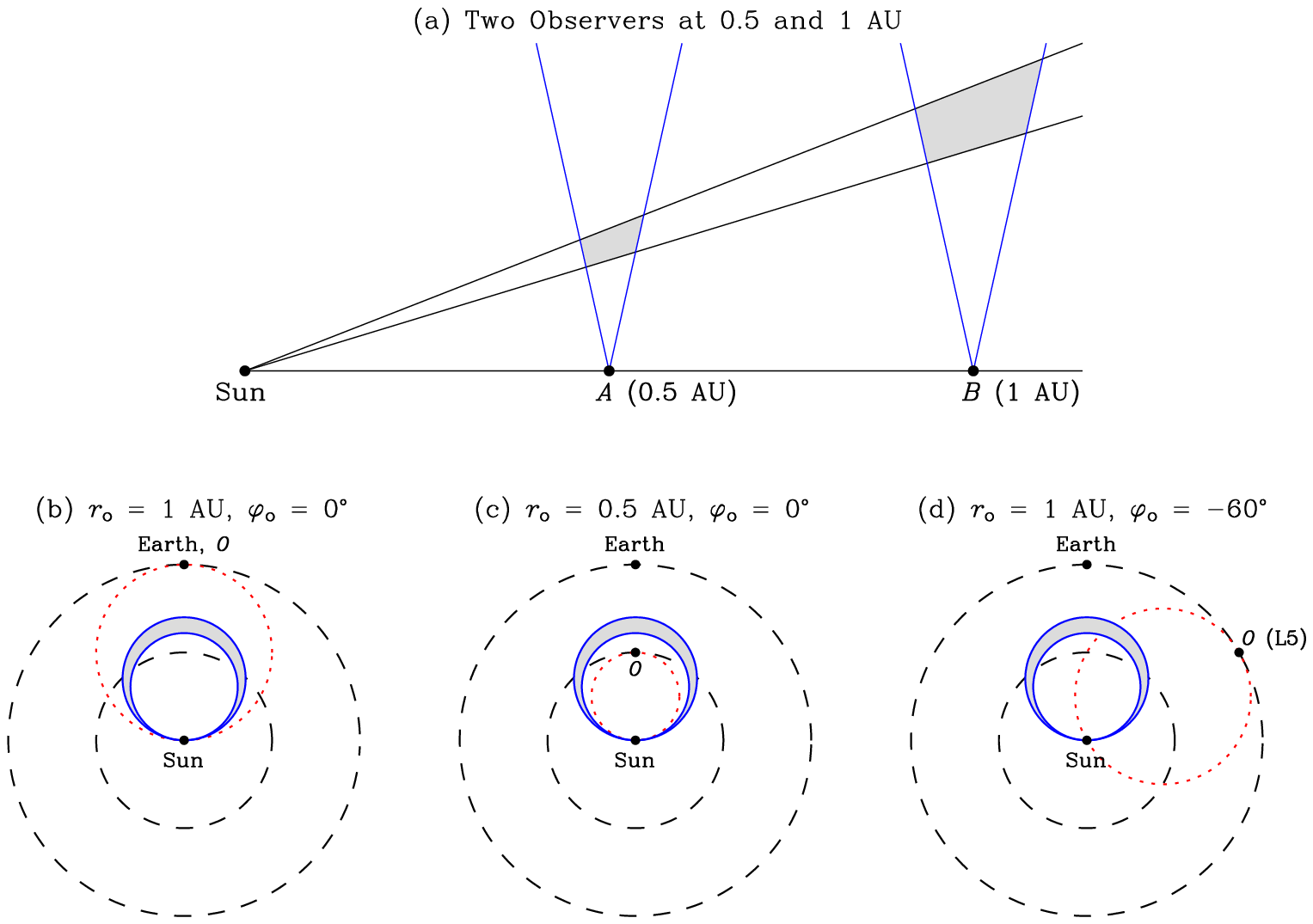}
\caption{Examples of the Thomson-scattering geometry for observers
at different radial distances $r_{\rm o}$ and longitudes
$\varphi_{\rm o}$. In panel (a), a radially-propagating solar wind
parcel is viewed sequentially, but at the same scattering angle,
by observers situated at radial distances of 0.5 AU (point $A$)
and 1 AU (point $B$). Panels (b--d) illustrate the observation of
an interplanetary sheath, denoted as a shaded region, by observers
with different combinations of $r_{\rm o}$ and $\varphi_{\rm o}$.
Longitude is defined to be positive (negative) for an observer
situated to the west (east) of Earth.} \label{00Cartoon}
\end{figure}

A plasma parcel emitted from the Sun would be observed, at the
same elongation and the same Thomson-scattering angle, firstly by
an observer situated at a radial distance of 0.5 AU from the Sun
centre, and subsequently by an observer at 1 AU (Figure
\ref{00Cartoon}a). Such a configuration was discussed
qualitatively by \citet{Jackson2010} and is analysed
quantitatively in Section \ref{Sec:HalfOneAU} of this paper.
Observations from {\it STEREO}/HI suggest that a travelling sheath
can be approximated as an expanding bubble
\citep[e.g.,][]{Howard-T-A2009a,Lugaz2010,Davies2012,Mostl2013}.
In-situ observations indicate that CMEs undergo self-similar
expansion, as the speed profiles within CMEs themselves tend to be
a linear function of time
\citep[e.g.,][]{Farrugia1993,Gulisano2012}. In the schematic
Figures \ref{00Cartoon}b-d, the sheath region following an
Earth-directed interplanetary shock is represented as a
self-similarly expanding bubble. The sheath can look quite
different when viewed from different heliocentric distances
(Figures \ref{00Cartoon}b and \ref{00Cartoon}c) and/or different
heliospheric longitudes (Figures \ref{00Cartoon}b and
\ref{00Cartoon}d).

\subsection{Thomson-Scattering WL Formulae} \label{Sec:method-wL}
A small parcel of free electrons, that is illuminated by a known
intensity of incident sunlight (measured in W m$^{-2}$), will
scatter a certain amount of power per unit solid angle (measured
in W rad$^{-1}$). The effect of the Thomson-scattering geometry
can be characterised by the so-called scattering angle $\chi$, as
depicted in Figure 1 of \citet{Xiong2013}. Scattering can be
backward ($\chi < 90^\circ$), perpendicular ($\chi = 90^\circ$),
and forward ($\chi
> 90^\circ$). All photons that are scattered into an optical cone defined by the
point spread function of an individual pixel will be attributed to
that pixel \citep[Figure 1b,][]{Xiong2013}. The classic principles
of WL Thomson-scattering, as applied to coronagraph observations
\citep{Billings1966}, have been adapted to heliospheric imaging
\citep{Vourlidas2006,Howard-T-A2009a,Jackson2010,Howard2012,Xiong2013}.
The transverse electric field oscillation $\delta \mathbf{E}$ of
the Thomson-scattered radiance, which is inherently a continuum,
can be considered in terms of its two orthogonal components, a
tangential component $\delta \mathbf{E}_{\rm T}$ and a radial
component $\delta \mathbf{E}_{\rm R}$. The amplitudes of these two
orthogonal oscillations ($I_{\rm T} = |\delta \mathbf{E}_{\rm
T}|^2$ and $I_{\rm R} = |\delta \mathbf{E}_{\rm R}|^2$) can be
measured separately, using a polariser. The total radiance $I$ and
degree of polarisation $p$ are defined as follows:
\begin{eqnarray}\label{Equ:Polarization}
 && I = I_{\rm T} + I_{\rm R}  \label{Equ:I}\\
 && p= \frac{I_{\rm T} - I_{\rm R}}{I}
\end{eqnarray}
Although the incident sunlight is unpolarised ($p=0$), the
scattered WL radiance remains unpolarised only when the scattering
angle $|\chi-90^\circ|=90^\circ$. The scattered light is
elliptically polarised ($0<p<1$) for $0^\circ <
|\chi-90^\circ|<90^\circ$ and linearly polarised ($p=1$) for $\chi
= 90^\circ$. Each pixel of a detector records the LOS integral of
local WL radiance.
\begin{equation}
 \begin{array}{l}
 \left( \begin{array}{c}
   I\\
   I_{\rm T}\\
   I_{\rm R}\\
 \end{array}\right) =
\displaystyle \int^\infty_0 \left( \begin{array}{c}
   i\\
   i_{\rm T}\\
   i_{\rm R}\\
 \end{array}\right) dz=
  \displaystyle \int^\infty_0 n \, z^2
\left( \begin{array}{c}
   G\\
   G_{\rm T}\\
   G_{\rm R}\\
 \end{array}\right) dz
 \end{array} \label{Equ:LOS}
\end{equation}
Here $z$ refers to a distance between the detector and the
scattering site, as shown in Figure 1b of \citet{Xiong2013}. The
mathematical expressions for $G$, $G_{\rm R}$, and $G_{\rm T}$ are
given by Equations 1 and 2 of \citet{Xiong2013}. The observed WL
radiance is determined jointly by the heliospheric distribution of
electrons $n$ and Thomson-scattering geometry factors ($z^2 G$,
$z^2 G_{\rm R}$, $z^2 G_{\rm T}$).

As noted above, the efficiency of Thomson scattering depends
significantly on the Thomson scattering angle $\chi$. The
perpendicular scattering, $\chi=90^\circ$, received by an observer
comes from the Thomson Sphere. The ``Thomson sphere", sometimes
called the ``Thomson surface", is the sphere in which the Sun and
observer lie at opposite ends of a diameter
\citep[e.g.,][]{Vourlidas2006,Howard2012}. The ecliptic cross
sections of the Thomson scattering spheres for three observers are
shown as dotted circles in Figures \ref{00Cartoon}b-d. The LOS
from an observer crosses its Thomson sphere at a so-called $p$
point \citep[Figure 2,][]{Tappin2004}, where both the intensity of
incident sunlight and local electron density are greatest, but the
efficiency of Thomson scattering is least. Competition between
these three effects results in the spread of local radiance ($i$,
$i_{\rm T}$, $i_{\rm R}$ in Equation \ref{Equ:LOS}) to large
distances from the Thomson surface, an effect that is greater at
larger elongations $\varepsilon$ from the Sun. \citet{Howard2012}
described this broad spreading effect, using the term ``Thomson
plateau". Namely, along a single LOS, the radiance per unit
electron density is virtually constant over a broad range of
scattering angles $\chi$ centred at the $p$ point. The Thomson
plateau, in terms of its relevance to heliospheric image, was
discussed in detail by \citet{Howard-T-A2009a},
\citet{Howard2012}, and \citet{Xiong2013}.

A major milestone in stereoscopic WL imaging of interplanetary
CMEs was achieved by the {\it STEREO}/HI instruments
\citep[e.g.,][]{Eyles2009,Davies2009,Davis2009,Harrison2009}. This
heliospheric imaging capability was built on the heritage of the
{\it Solar Mass Ejection Imager} ({\it SMEI}) instrument on the
{\it Coriolis} spacecraft \citep{Eyles2003}. The {\it STEREO}
mission is comprised of two spacecraft, with one leading ({\it
STEREO} A) and the other trailing ({\it STEREO} B) the Earth in
its orbit. Both spacecraft separate from the Earth by $22.5^\circ$
per year. The HI instrument on each {\it STEREO} spacecraft
consists of two cameras, HI-1 and HI-2, whose optical axes lie in
the ecliptic. Elongation coverage in the ecliptic is $4^\circ$ --
$24^\circ$ for HI-1 and $18.7^\circ$ -- $88.7^\circ$ for HI-2. The
field of view (FOV) is $20^\circ \times 20^\circ$ for HI-1 and
$70^\circ \times 70^\circ$ for HI-2. The cadence of HI-1 is
usually 40 minutes and that of HI-2 is 2 hours \citep{Eyles2009}.
The current generation of heliospheric imagers do not have WL
polarisers. Polarisation measurements have, up until now, only
been made by coronagraphs
\citep[e.g.,][]{Poland1976,Crifo1983,Moran2004,Pizzo2004,deKoning2009,Moran2010}.
For instance, \citet{Moran2004} used polarisation measurements of
WL radiance by the {\it SOHO}/LASCO coronagraph to reconstruct CME
orientations near the Sun.

Sky maps, often presented in the Hammer-Aitoff projection, can be
used to highlight and track WL transients
\citep[e.g.,][]{Tappin2004,Zhang2013}. Time-elongation maps
(J-maps) are usually constructed by stacking differenced radiance
between observed sky maps along a fixed position angle (sometimes
background subtracted images are used instead of difference
images). Using such J-maps, transients such as CMEs are manifest
as inclined streaks
\citep[e.g.,][]{Sheeley2008,Rouillard2008,Xiong2011,Harrison2012,Davies2012,Xiong2013}.
As a propagating transient is viewed along larger elongations, its
WL signatures become fainter.

\subsection{Faraday Rotation Formula} \label{Sec:method-FR}
Due to a FR effect, the plane of polarisation of linearly
polarised radio emission is continuously rotated as the radio wave
passes through the heliosphere. For radio waves, the ubiquitous
magnetised solar wind flow serves as a magneto-optical
birefringence medium. The formulae for FR are expressed below:
\begin{eqnarray}
  && \Omega = \Omega_{\rm RM} \cdot \lambda^2  \label{Equ:FR1} \\
  && \Omega_{\rm RM} = \int \omega_{\rm RM} \, dz  \label{Equ:FR2} \\
  && \omega_{\rm RM} = \frac{e^3}{8 \, \pi^2 \, \epsilon_0 \, m_{\rm e}^2 \, c^3} \, n \, B_\parallel  \nonumber\\
  &&   \quad\quad \, = \left [\, 2.63 \times 10^{-13} \,\, \frac{\mbox{rad}}{\mbox{T}} \, \right ] \,\, n \, B_\parallel \label{Equ:FR3} \\
  && \delta \, \omega_{\rm RM} \propto \delta\,(n\,B_\parallel) \\
  && \delta \, \Omega_{\rm RM} \propto \int \delta \,\omega_{\rm RM} \,
  dz  \label{Equ:FR5}
\end{eqnarray}
Where $q$, $\epsilon$, $m_{\rm e}$, and $c$ represent the
constants that are the electron charge, the permittivity of free
space, the mass of an electron, and the speed of light,
respectively. A FR measurement of $\Omega_{\rm RM}=1$ rad m$^{-2}$
corresponds to $\Omega=0.97^\circ$ at 2.3 GHz (wavelength
$\lambda=0.13$ m), $\Omega=57.3^\circ$ at 300 MHz ($\lambda=1$ m),
and $\Omega=1432^\circ$ at 60 MHz ($\lambda=5$ m). The calibration
of ground-based FR observations is difficult, as the radio wave
passes through the magnetised plasma of the ionosphere,
magnetosphere (including the plasmasphere), and solar wind.
\citet{Oberoi2012} surveyed and compared the FR signatures
associated with each of these different regions.

A large portion of the inner heliosphere can be monitored, using
FR imaging. Prime heliospheric targets measured in FR include
interplanetary CMEs and CIRs \citep{Oberoi2012}. Because the
low-frequency radio interferometers such as the {\it MWA}, {\it
LOFAR}, and {\it VLA} feature a wide FOV, high sensitivity, and
multi-beam forming capabilities, it is expected to be capable of
mapping the magnetic field in the inner heliosphere with a
remarkable sensitivity. The high sensitivity of FR measurements
enables fluctuations in the heliospheric/interstellar magnetic
field and plasma density, resulting from MHD turbulence, to be
revealed \citep[e.g.,][]{Jokipii1969,Goldshmidt1993,Hollweg2010}.
For instance, gradients in FR measurement have been observed
across active Galactic Nuclei (AGN) jets, using the {\it Very Long
Baseline Array}, which demonstrate that ordered helical magnetic
fields are associated with these jets
\citep[e.g.,][]{Zavala2002,Gomez2008,Reichstein2012}. The sheath
region associated with a fast CME can be similarly probed. FR
measurements of the sheath would provide a value for $\Omega_{\rm
RM}$ in Equations \ref{Equ:FR1}-\ref{Equ:FR5}. Any measured value
of the FR, $\Omega_{\rm RM}$, would correspond to a statistical
average, as the plasma and magnetic fields within such sheath
regions are in a highly turbulent state.

\section{White-Light and Faraday Rotation Signals Received at 0.5 and 1 AU} \label{Sec:HalfOneAU}
The remote imaging in WL and FR of an Earth-directed sheath from
two vantage points, one at 0.5 AU and the other at 1 AU, is
considered in Section \ref{Sec:HalfOneAU-compare}. Section
\ref{Sec:HalfOneAU-WL} demonstrates how spatial position and
electron number density can be inferred from polarisation
observations of WL radiance. Section \ref{Sec:HalfOneAU-FR}
presents a means by which magnetic field can be diagnosed from FR
measurements.

\subsection{Comparing Remotely-Sensed WL and FR Observations from Different Vantage Points} \label{Sec:HalfOneAU-compare}
\begin{figure}
  \includegraphics{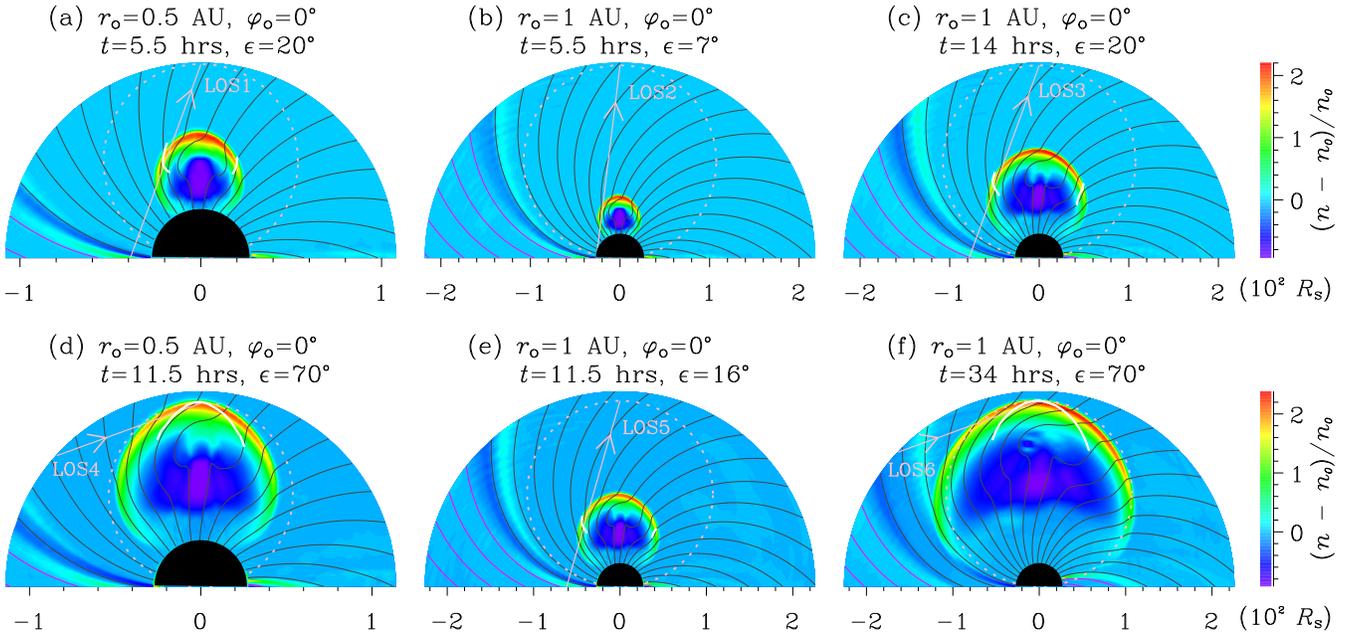}
\caption{Relative enhancement of electron number density, in the
ecliptic, $(n-n_0)/n_0$ within an Earth-directed interplanetary
sheath. Red and black solid lines indicate sunward and
anti-sunward interplanetary magnetic field lines, respectively.
The sheath is imaged by two observers on the Sun-Earth line
($\varphi_{\rm}=0^\circ$), at heliocentric distances of $r_{\rm
o}=0.5$ (left column) and 1 AU (central and right columns). For
each observer, the ecliptic cross-section of its corresponding WL
Thomson-scattering sphere is depicted as a dotted circle. Six
lines-of-sight, LOS1--LOS6, are superimposed as straight arrows.
All LOSs look westward. At any given time $t$, the two observers,
both located on the Sun-Earth line, detect the sheath at different
elongations $\varepsilon$ (compare panel a with panel b, and panel
d with panel e). Conversely, when viewing along the same
$\varepsilon$, the two observers detect the sheath at different
$t$ (compare panel a with panel c, and panel d with panel f).
Solid white curves overlaid on each panel indicate the position of
the sheath inferred from polarised WL imaging.} \label{10Contour}
\end{figure}

Figure \ref{10Contour} shows the modelling results of an
Earth-directed sheath propagating from the Sun to 1 AU. The
travelling sheath is supposed to be imaged simultaneously by two
observers at 0.5 and 1 AU. The WL and FR signatures of the sheath
are synthesised, using the methods in Section \ref{Sec:method}.
Representative LOSs, which cut through the sheath (LOS1--6), are
denoted using arrows in Figure \ref{10Contour}. The variations of
various physical parameters along LOS1--3 are shown in Figure
\ref{20LOS-a}. LOS1, LOS2, LOS3, and LOS5 are approximately
tangential to the left flank of the shock; LOS4 and LOS6 are
tangential to the nose of the shock. LOS1 and LOS4 are directed
towards the observer situated at 0.5 AU; all other LOSs are
directed towards the observer at 1 AU. The viewing configuration
for LOS1 (Figure \ref{10Contour}a) is equivalent to that for LOS3
(Figure \ref{10Contour}c), as the elongation of the shock front is
the same for LOS1 and LOS3. Thus, the Thomson scattering geometry
is identical for these two LOSs, leading to similar LOS profiles
in Figure \ref{20LOS-a}. Of course, the observed radiance along
LOS1 is much stronger than that along LOS3 (Table \ref{Tab:LOS}).
Similarly, the observations along LOS4 and LOS6 (Figures
\ref{10Contour}d and \ref{10Contour}f) have identical Thomson
scattering geometries. At any given time, the sheath is viewed at
greater elongations from a vantage point closer to the Sun. For
instance, at an elapsed time of 5.5 hours, the foremost elongation
$\varepsilon$ of the sheath is $20^\circ$ for an observer at 1 AU
(LOS1) compared with $7^\circ$ for an observer at 0.5 AU (LOS2).
While the sheath is undetectable in WL along LOS2 (Figure
\ref{20LOS-a}g), it can be observed in FR (Figure \ref{20LOS-a}l).
The portion of an LOS that contributes most to the WL radiance
broadens and flattens with increasing elongation, and shifts
gradually towards the observer. At elongations beyond $90^\circ$,
only back-scattered photons are received; electrons in the
vicinity of the observer mainly contribute to remote-sensing
signatures for elongations beyond $90^\circ$. Such observations
for elongations of $\varepsilon \ge 90^\circ$ are less useful for
the purposes of space weather prediction. In Figures
\ref{10Contour}d and \ref{10Contour}f, the shock front has already
reached the observer, and can be detected in-situ.

\subsection{Inferences of Sheath Position from Polarised White Light} \label{Sec:HalfOneAU-WL}
\begin{figure}
  \includegraphics[width=.95\textwidth]{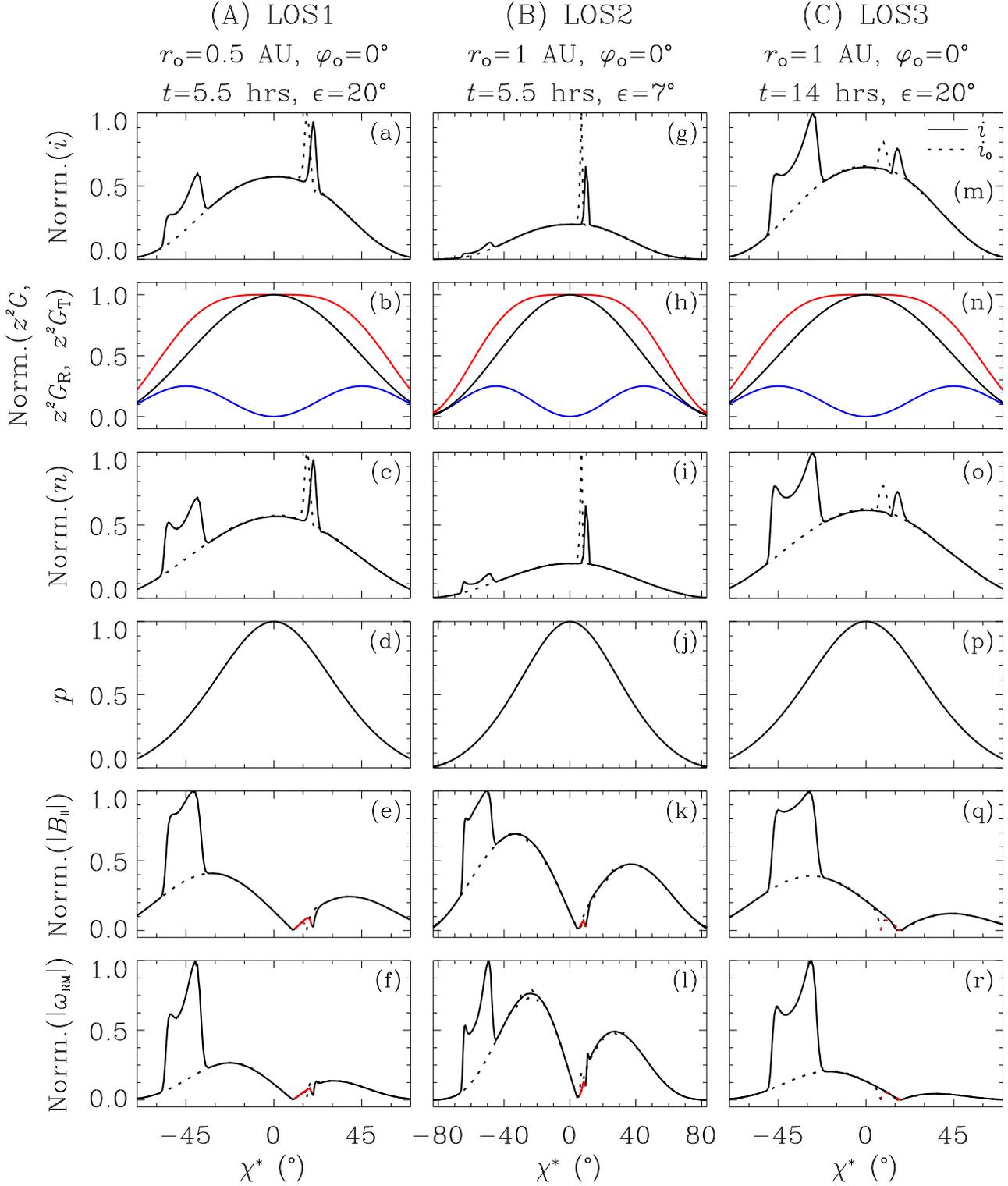}
\caption{WL radiance $i$, WL Thomson-scattering geometry factors
($z^2 G$, $z^2 G_{\rm R}$, $z^2 G_{\rm T}$), electron number
density $n$, degree of WL polarisation $p$, parallel magnetic
field $|B_\parallel|$, and FR measurement $|\omega_{\rm RM}|$,
plotted as a function of modified scattering angle
$\chi^*=90-\chi$ along LOS1 (column A), LOS2 (column B), and LOS3
(column C). Each parameter is normalised to its maximum value
along each LOS, as given in Table \ref{Tab:LOS}. Note that $i=n
\,Z^2 G$ and $\omega_{\rm RM} \propto n \, B_\parallel$. For the
parameters $i$, $n$, $|B_\parallel|$, and $|\omega_{\rm RM}|$,
initial and disturbed profiles are depicted as dashed and solid
lines, respectively. The black and red sections of the
$|B_\parallel|$ and $|\omega_{\rm RM}|$ profiles (two bottom rows)
correspond to negative and positive values of $B_\parallel$,
respectively.} \label{20LOS-a}
\end{figure}

\begin{figure}
  \includegraphics[width=.45\textwidth]{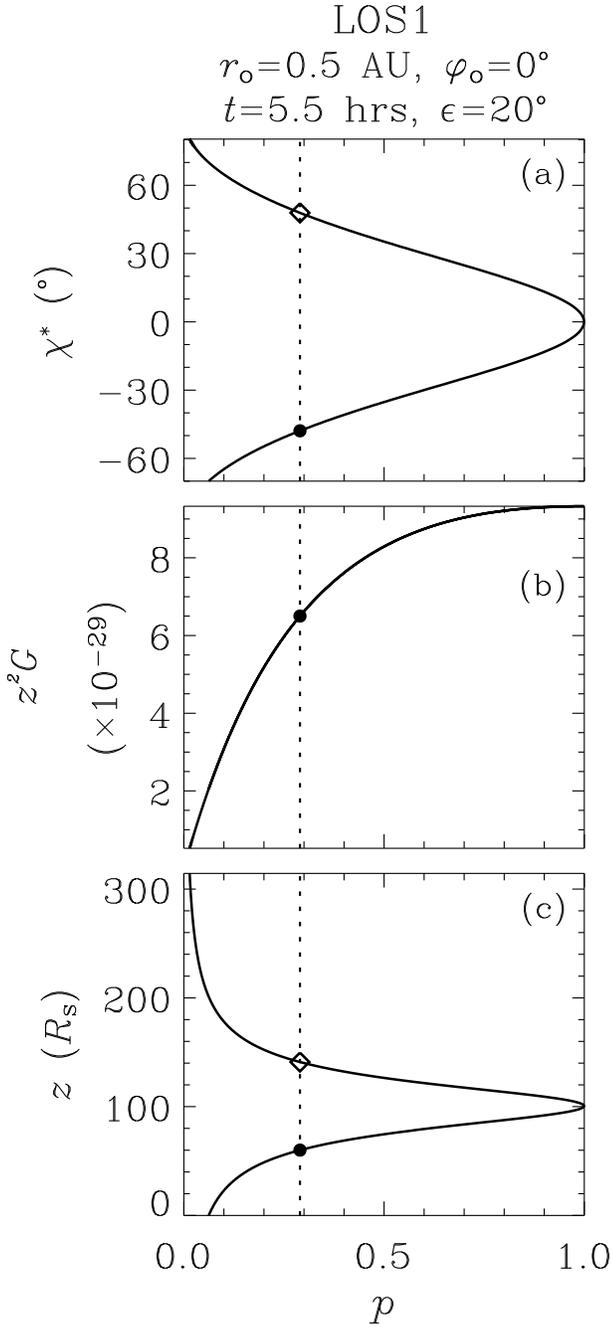}
\caption{Dependence of the scattering angle $\chi^*$, the
Thomson-scattering geometry factor $z^2 G$, and the LOS depth $z$
on the degree of polarisation $p$ for LOS1 in Figure
\ref{10Contour}a. In response to shock passage, the initial
radiance components, $I_{\rm T0}$ and $I_{\rm R0}$, are enhanced
to values of $I_{\rm T}$ and $I_{\rm R}$, respectively. The
enhancement in these radiance components determines the modified
degree of polarisation $p^*$, according to the expression
$p^*=\frac{I_{\rm T}- I_{\rm T0} - I_{\rm R}+ I_{\rm R0}}{I -
I_0}$. The vertical dashed line, demarking $p^*$, crosses the
$\chi^*$ and $z$ profiles twice.} \label{30LOS-1a}
\end{figure}

The WL radiance of CMEs is determined by both the electron number
density distribution and the Thomson-scattering geometry (Equation
\ref{Equ:LOS}). The total radiance at a scattering site ($i$), and
its constituent radial ($i_{\rm R}$) and tangential ($i_{\rm T}$)
components, are associated with Thomson-scattering factors $z^2 \,
G$, $z^2 \, G_{\rm R}$, and $z^2 \, G_{\rm T}$, respectively. Near
the Thomson-scattering surface, $z^2 \, G_{\rm T}$ is much larger
than $z^2 \, G_{\rm R}$. If a dense parcel of plasma, viewed at
large elongations, approaches the Thomson surface, its WL
signatures will comprise (1) an increase in $I$, (2) an increase
in $I_{\rm T}$, (3) a decrease in $I_{\rm R}$, and (4) an increase
in the degree of polarisation $p$. The variation of $p$ is
largest, while that of $I$ is negligible. A plasma parcel's
distance from the Thomson sphere has a less significant effect on
$I$ at larger elongations. However, the determination of the
plasma parcel's location will be more uncertain, if only
unpolarised WL observations are available, as with current
operational heliospheric imaging systems. Polarisation
observations can provide an important clue to the primary
scattering site. LOS1 in Figure \ref{10Contour} is used to
demonstrate these inferences. The Thomson-scattering geometry is
independent of the distribution of heliospheric electrons. The
degree of polarisation ($p$) and the Thomson-scattering factors
($z^2 \, G$, $z^2 \, G_{\rm R}$, $z^2 \, G_{\rm T}$), as presented
in Figure \ref{20LOS-a}b and \ref{20LOS-a}d, only depend on the
modified scattering angle $\chi^*=90^\circ - \chi$. The profiles
of $p$, $z^2 \, G$, $z^2 \, G_{\rm R}$, and $z^2 \, G_{\rm T}$ are
symmetrical around $\chi^*=0^\circ$. The dependence of $\chi^*$,
$z^2 \, G$, and LOS distance $z$ on $p$ can be seen in Figure
\ref{30LOS-1a}. $p=1$ corresponds to perpendicular scattering
(i.e. $\chi^*=0^\circ$). $p \ne 1$ corresponds to two solutions
for $\chi^*$: one resulting from forward scattering ($\chi^* <
0^\circ$), and the other associated with backward scattering
($\chi^* > 0^\circ$). In response to the passage of the shock, the
initial radiance components at $t=0$, $I_{\rm T0}$ and $I_{\rm
R0}$, are enhanced to values denoted by $I_{\rm T}$ and $I_{\rm
R}$, respectively. The increase in the radiance components define
a so-called modified degree of polarisation that we denote using
$p^*$. $p^*$ is given by
\begin{equation}\label{Equ:p*}
   p^*=\frac{I_{\rm T}- I_{\rm T0} - I_{\rm
R}+ I_{\rm R0}}{I - I_0}
\end{equation}
$p$ is 0.62 along LOS1 at $t=0$ hours. This effectively defines
the degree of polarisation associated with the background solar
wind. During the sheath passage, at $t=5.5$ hours, $p$ is 0.58
along this LOS. The modified degree of polarisation $p^*$, derived
using Equation \ref{Equ:p*}, is therefore 0.29 at $t=5.5$ hours.
The radiance enhancement is due to the presence of the sheath in
the LOS. The sheath, which trails the shock front, occupies a
relatively small volume of interplanetary space. The sheath
occupies the portion of LOS1 bounded by $-55^\circ < \chi^* <
-35^\circ$ (Figures \ref{20LOS-a}a and \ref{20LOS-a}c). Within
this region, $p$ smoothly varies from 0.15 to 0.5 (Figure
\ref{20LOS-a}d). The average value of $p^*$ within the sheath is
0.29. In an inverse approach, $p^*$ can be used to estimate the
scattering angle $\chi^*$ within the sheath. This is demonstrated
in Figure \ref{30LOS-1a}a. $p^*=0.29$ corresponds to $\chi^*=\pm
\, 46^\circ$ and $\overline{z^2 \, G} = 6.5 \times 10^{-29}$,
where $\overline{z^2 G}$ is the average value of $\ {z^2 G}$ in
the sheath. The solution of $\chi^*= 46^\circ$ can be immediately
excluded, as an Earth-directed CME can generally be identified
(indeed much earlier) as being front-sided based on Extreme
Ultraviolet (EUV) images of the full solar disk
\citep[e.g.,][]{Thompson1998,Plunkett1998}. The other solution,
$\chi^*= - 46^\circ$, is physical and yields a value of $60\,
R_{\rm S}$ for the distance, $z$, of the main scattering site
(corresponding to the sheath) from the detector. How best to judge
which solutions for $\chi^*$ are physical is explained in detail
in Section \ref{Sec:Pattern}. Once the Thomson-scattering factor
$z^2 \, G$ of the sheath has been inferred, its column-integrated
electron number density can be estimated based on the following
equation:
\begin{equation}
  \delta N_{\rm sheath} = \int \delta n \, dz \simeq \frac{\int \delta n \, z^2 \, G \, dz}{\overline{z^2 \, G}}
\end{equation}
It is clear that WL polarisation measurement can prove extremely
valuable in the study of interplanetary CMEs and shocks.

\subsection{Magnetic Field Inferred from Faraday Rotation} \label{Sec:HalfOneAU-FR}
\begin{figure}
  \includegraphics[width=.45\textwidth]{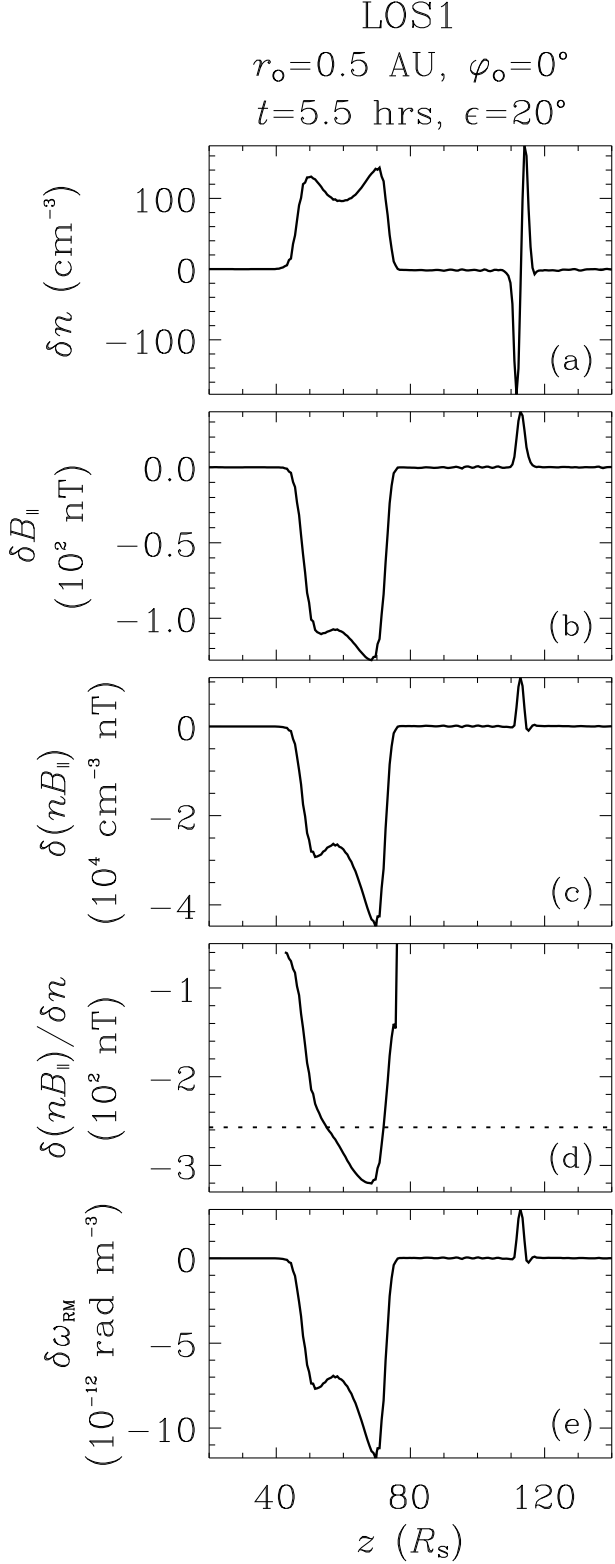}
\caption{The deviation of various parameters, from their initial
values, plotted as a function of $z$ along LOS1. Note that
$\frac{\delta\, (n \, B_\parallel)}{\delta \, n}= \delta
B_\parallel + B_{\parallel 0} + n_0 \frac{\delta
B_\parallel}{\delta n} = B_\parallel + n_0 \frac{\delta
B_\parallel}{\delta n}$. The parallel magnetic field component
$B_\parallel^*$ (plotted as a horizontal dashed line in panel d)
is calculated using the expression $\frac{\delta \, \Omega_{\rm
RM}}{\delta \, I} =\frac{\int \delta \, \omega_{\rm RM} \,\,
dz}{\int \delta \, i \,\, dz} = \frac{\int \delta \, (n \,
B_\parallel) \,\, dz}{\int z^2 G \, \delta n \,\, dz} \approx
\frac{1}{\,\,\overline{z^2 G}\,\,} \, \frac{\delta\, (n \,
B_\parallel)}{\delta \, n} = \frac{1}{\,\,\overline{z^2 G}\,\,} \,
B_\parallel^*$.} \label{31LOS-1b}
\end{figure}

As discussed in Section \ref{Sec:HalfOneAU-WL}, the
column-integrated electron number density along any LOS can be
inferred from its WL observations. Thus, if a radio beam lies
within the FOV of a WL imager such that they remotely probe the
same plasma volume, the WL density measurements can be used to
retrieve magnetic field strength from the received FR signal. We
demonstrate this, for LOS1, in Figure \ref{31LOS-1b}. After
subtracting the background solar wind contribution, the
enhancements in FR measurement and WL radiance, due to the
presence of the sheath of the simulated Earth-directed shock, are
given by $\delta \, \Omega_{\rm RM}$ and $\delta I$, respectively.
The ratio of $\delta \, \Omega_{\rm RM}$ and $\delta I$ can be
expressed as
\begin{eqnarray}\label{Equ:inferB}
 \frac{\delta \, \Omega_{\rm RM}}{\delta \, I}  && =\frac{\int \delta \, \omega_{\rm RM} \,\, dz}{\int \delta \, i \,\, dz}
= \frac{\int \delta \, (n \, B_\parallel) \,\, dz}{\int z^2 G \,
\delta n \,\, dz} \nonumber\\
 && \approx \frac{\int \delta \, (n \, B_\parallel) \,\,
dz}{\,\overline{z^2 G} \, \int \delta n \,\, dz} \approx
\frac{1}{\,\,\overline{z^2 G}\,\,} \, \frac{\delta\, (n \,
B_\parallel)}{\delta \, n}
\end{eqnarray}
As discussed in Section \ref{Sec:HalfOneAU-WL}, $\overline{z^2 G}$
corresponds to the average value of $z^2 G$ in the sheath. The
derivable parameter $\frac{\delta\, (n \,B_\parallel)}{\delta \,
n}$, which we call $B_\parallel^*$, can be expressed in the form
\begin{equation}\label{Equ:B*}
   B_\parallel^* \equiv \frac{\delta\, (n \, B_\parallel)}{\delta \, n}= \delta B_\parallel +
 B_{\parallel 0} + n_0 \frac{\delta B_\parallel}{\delta n} = B_\parallel + n_0 \frac{\delta B_\parallel}{\delta n}
  > B_\parallel
\end{equation}
where $B_{\parallel 0}$ and $n_0$ denote the initial background
values of $B_{\parallel}$ and $n$, respectively. The inferred
value of $B_\parallel^*$ serves as an upper limit for
$B_\parallel$.

%------------------------------------------------------------------------------------------------------------------
\section{Radiance Patterns in J-Maps of White Light and Faraday Rotation} \label{Sec:Pattern}
\begin{figure}
  \includegraphics[width=.8\textwidth]{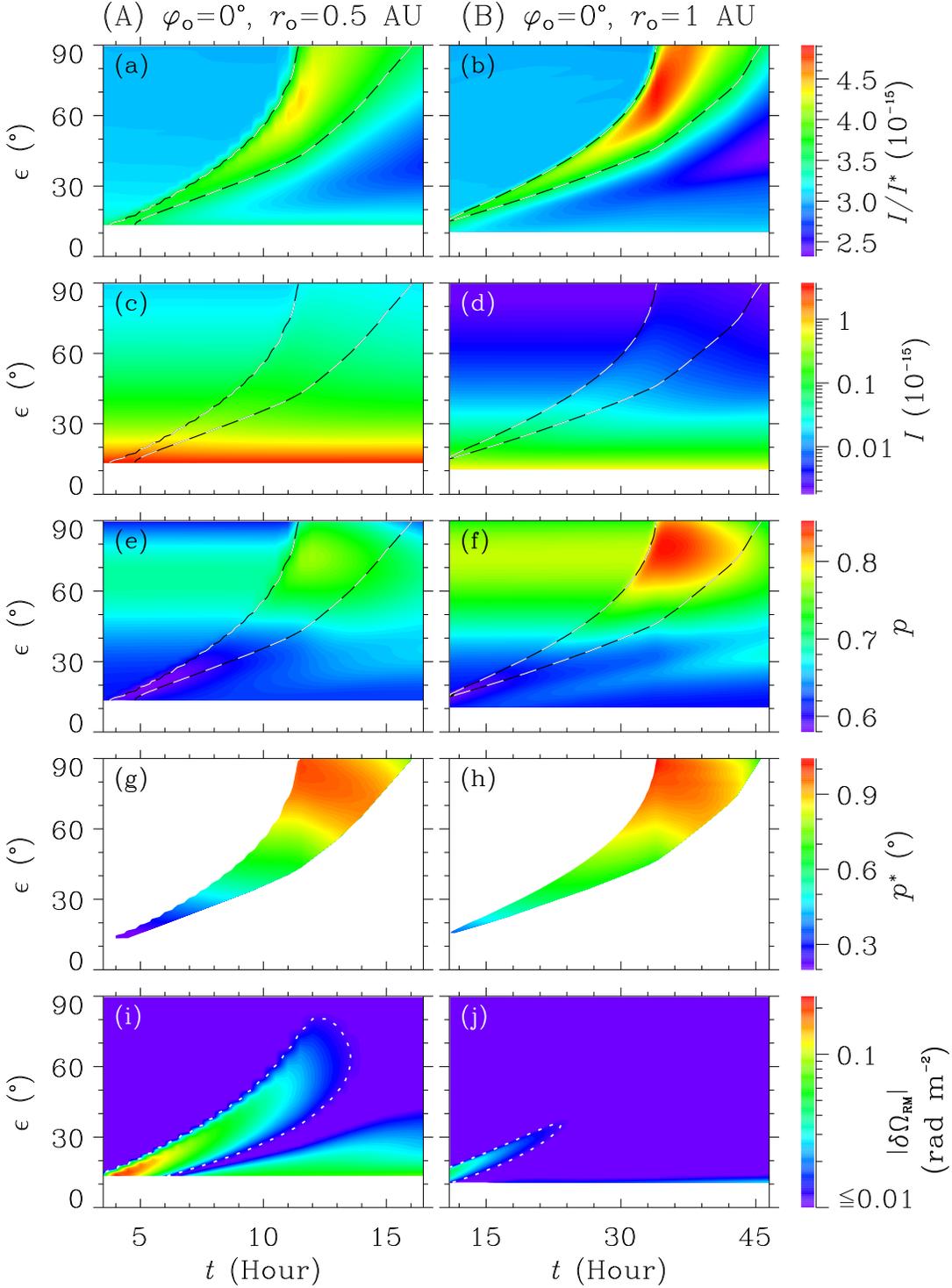}
\caption{Time-elongation maps of WL radiance (panels a and b:
$I/I^*$, panels c and d: $I$), degree of WL polarisation (panels e
and f: $p$, panels g and h: $p^*$), and FR measurement (panels i
and j: $|\delta \, \Omega_{\rm RM}|$), as viewed by observers at
longitude $\varphi_{\rm o}=0^\circ$ and at radii $r_{\rm o}=$ 0.5
AU (left column) and 1 AU (right column). $I^*$ is the
normalisation factor for $I$, and corresponds to the radial
variation of electron number density $n \propto r^{-2}$. The
dotted lines in panels a--f correspond to $I /I^* = 3.68 \times
10^{-15}$. $p$ and $p^*$ are determined from the radiance and the
enhancement in the radiance, respectively, as given by
$p=\frac{I_{\rm T} - I_{\rm R}}{I}$ and $p^*=\frac{I_{\rm T}-
I_{\rm T0} - I_{\rm R}+ I_{\rm R0}}{I - I_0}$. Panels g and h only
show $p^*$ within the sheath region.} \label{40Contour_Obs}
\end{figure}

\begin{figure}
 \includegraphics[width=.45\textwidth]{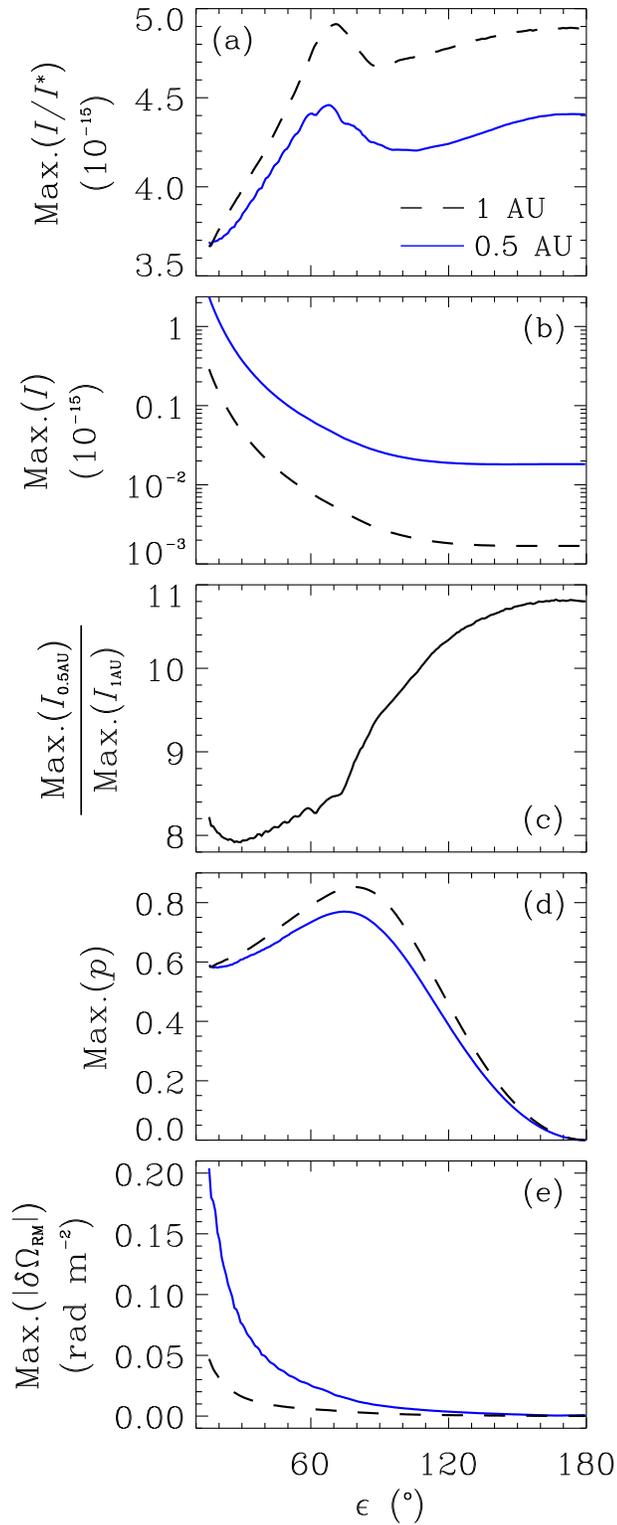}
\caption{A comparison of the WL radiance $I$ (panels a, b, and c),
the degree of WL polarisation $p$ (panel d), and the FR
measurement $\delta \, \Omega_{\rm RM}$ (panel e) over the
elongation range from $15^\circ$ to $180^\circ$, as viewed by
observers at 0.5 AU (solid line) and 1 AU (dashed line). The
attribution ``Max" refers to the strongest signal during the
sheath passage.} \label{50Compare}
\end{figure}

Shock propagation through the inner heliosphere can be identified
through the inclined trace with which it is associated in a
time-elongation map (J-map). J-maps of WL radiance ($I/I^*$, $I$)
and degree of polarisation ($p$, $p^*$), and FR measurement
$|\delta \, \Omega_{\rm RM}|$, as viewed from observers at 0.5 and
1 AU, are presented in Figure \ref{40Contour_Obs} and compared in
Figure \ref{50Compare}. The normalisation factor $I^*$ in Figure
\ref{40Contour_Obs} corresponds to an electron number density
distribution that varies according to $n \propto r^{-2}$. A
radiance threshold of $I/I^* \ge 3.68 \times 10^{-15}$ is used to
demarcate the sheath region in time-elongation ($t - \varepsilon$)
parameter space. The modified polarisation $p^*$ is only
calculated, using Equation \ref{Equ:p*}, inside the sheath
(Figures \ref{40Contour_Obs}g-h). The absolute values of $I$ and
$|\delta \, \Omega_{\rm RM}|$ within the sheath region are much
larger for the observer at 0.5 AU, whereas the sheath values of
$I/I^*$, $p$, and $p^*$ are comparable when viewed from either
vantage point. Over the elongation range $15^\circ \le \varepsilon
\le 180^\circ$, the radiance ratio $\frac{\mbox{Max.}(I_{\rm 0.5
AU})}{\mbox{Max.}(I_{\rm 1 AU})}$ is limited to values between 8
and 11 (Figure \ref{50Compare}c). This demonstrates that
interplanetary CMEs and shocks are viewed better from a location
closer to the Sun.

\begin{figure}
 \includegraphics[width=.99\textwidth]{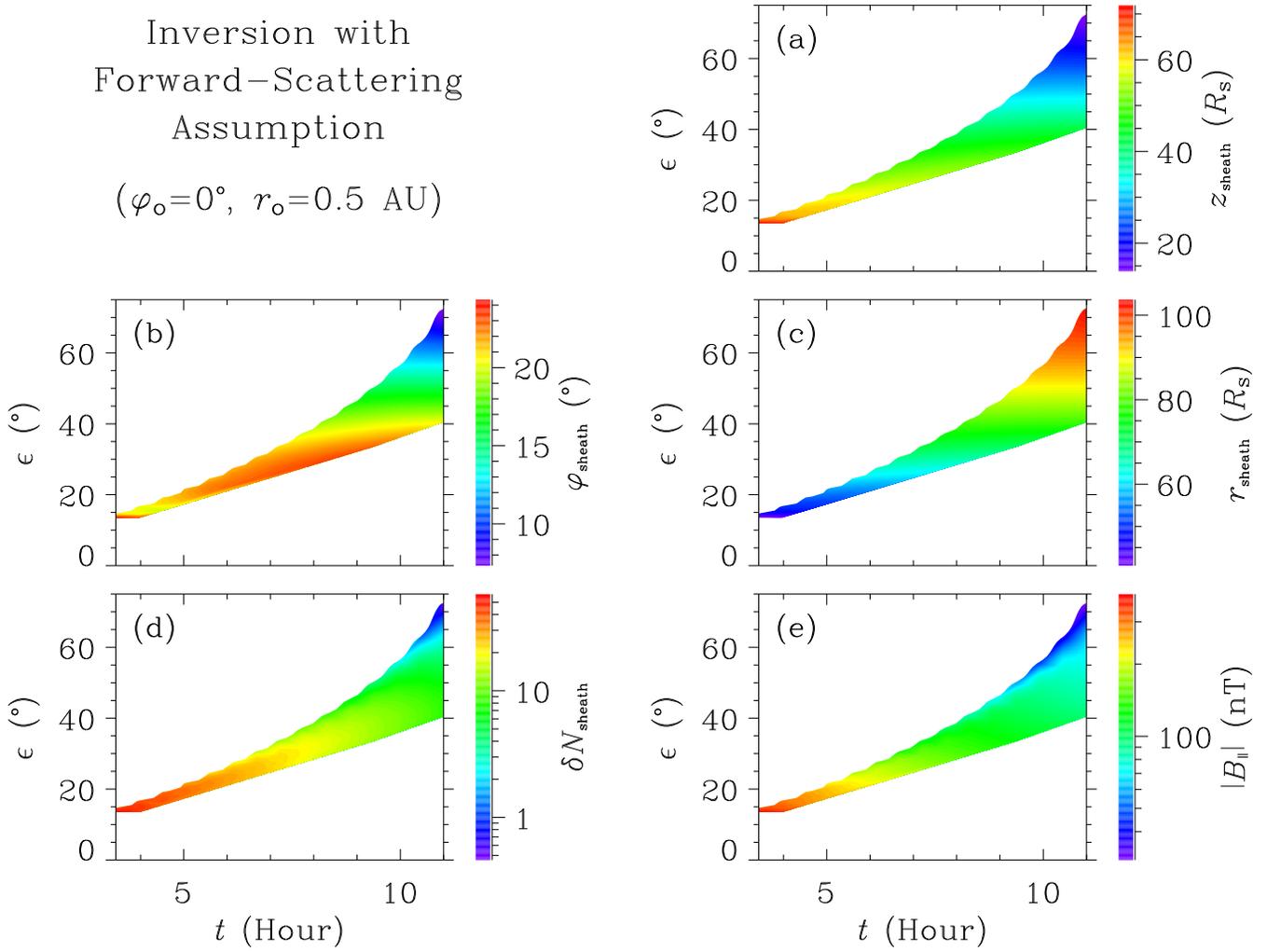}
\caption{Panels a--e present the location of the scattering site
($z_{\rm sheath}$, $\varphi_{\rm sheath}$, $r_{\rm sheath}$), the
column-integrated electron number density $\delta N_{\rm sheath}$,
and the parallel magnetic field $B_\parallel$, plotted as a
function of time and elongation, derived by assuming
forward-scattering ($\chi^* < 0^\circ$).} \label{41Contour_Inv1}
\end{figure}

\begin{figure}
  \includegraphics[width=.99\textwidth]{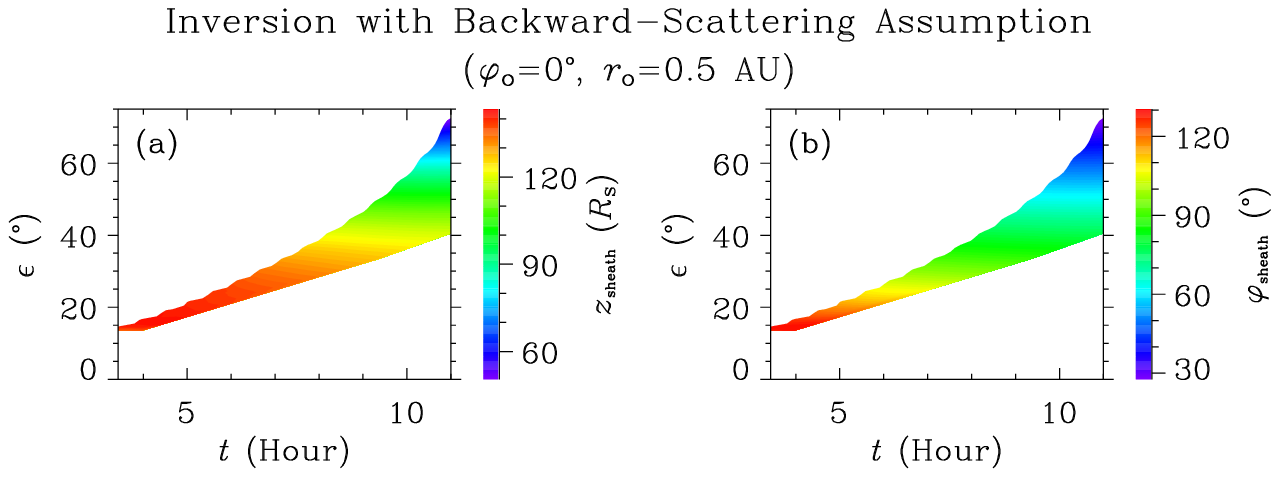}
\caption{Panels a--b present the location of the scattering site
($z_{\rm sheath}$, $\varphi_{\rm sheath}$), plotted as a function
of time and elongation, derived by assuming backward-scattering
($\chi^* > 0^\circ$).} \label{42Contour_Inv2}
\end{figure}

The position, mass, and magnetic field of the sheath can be
inferred from those directly-measurable parameters presented in
Figure \ref{40Contour_Obs}, using the analytical methods presented
in Sections \ref{Sec:method-wL} and \ref{Sec:HalfOneAU-FR}. As
shown in Figure \ref{30LOS-1a}, and explained in Section
\ref{Sec:HalfOneAU-WL}, the Thomson-scattering factors are
symmetrical around $\chi^*=0^\circ$. As a result, a single value
of degree of polarisation $p^*$ corresponds to two symmetrical
solutions for the scattering angle $\chi^*$. The results shown in
Figure \ref{41Contour_Inv1} are derived from those in Figure
\ref{40Contour_Obs} (for an observer at 0.5 AU) under the
assumption of forward scattering, while those in Figure
\ref{42Contour_Inv2} assume backward scattering. For the
forward-scattering situation, presented in Figure
\ref{41Contour_Inv1}b, the inferred longitude of the sheath
$\varphi_{\rm sheath}$ is $22^\circ$ at an elapsed time of 5 hours
and $9^\circ$ at 11 hours. For the backward-scattering case, shown
in Figure \ref{42Contour_Inv2}b, the sheath is at $\varphi_{\rm
sheath}=140^\circ$ and $60^\circ$ at these times. So, an observer
at 0.5 AU infers a longitude change $\Delta \varphi_{\rm sheath}$
of $13^\circ$ (Figure \ref{41Contour_Inv1}b) and $80^\circ$
(Figure \ref{42Contour_Inv2}b) for the forward and
backward-scattering cases, respectively. The dramatic change in
sheath longitude for the backward-scattering case might indicate
that the shock is significantly deflected during its
interplanetary propagation. However, such an abnormal degree of
lateral deflection of $\Delta \varphi_{\rm sheath}=80^\circ$ would
be highly unphysical, and may imply a ``ghost trajectory"
\citep[Figure 6,][]{DeForest2013}. The east-west symmetry of the
radiance pattern suggests that the shock is actually front-sided
and Earth-directed, rendering the assumption of
backward-scattering invalid (Figure \ref{42Contour_Inv2}). If we
assume that the radiance pattern shown in Figure
\ref{40Contour_Obs} is attributable to forward scattering, the
inferred position of the sheath is shown as the solid white curve
in Figure \ref{10Contour}. This agrees very well with the actual
position of the sheath. At any given time, only a certain portion
of the sheath will be visible from a fixed observing location
\citep{Xiong2013}. For example, at an elapsed time of 5.5 hours,
it is the flank of the sheath (Figure \ref{10Contour}a) that
corresponds to the leading edge of the radiance pattern in
\ref{40Contour_Obs}a, while 6 hours later it is the nose (Figure
\ref{10Contour}d). So, in fact, the longitudinal change of $\Delta
\varphi_{\rm sheath}=13^\circ$ inferred from Figure
\ref{41Contour_Inv1}b is actually an artefact of the viewing
geometry and does not represent an actual deflection of the shock
front. Along with the inferred position of the sheath, the
column-integrated electron number density, $\delta N_{\rm
sheath}$, and the parallel magnetic field component,
$B_\parallel$, are also presented in Figure \ref{41Contour_Inv1}.
The derived value of $|B_\parallel|$ provides an upper limit for
the actual magnetic field, as explained in Section
\ref{Sec:HalfOneAU-FR}. By making coordinated observations in WL
and FR, CMEs can not only be continuously tracked, but
quantitatively diagnosed as they propagate through interplanetary
space.

\section{Interplanetary Imaging from Different Observation Sites} \label{Sec:Depend}
An interplanetary CME looks different when viewed from different
vantage points, but can be readily imaged from a wide range of
longitudes. The observed WL radiance pattern depends not only on
the longitude $\varphi_{\rm o}$ of the observer, as discussed by
\citet{Xiong2013}, but also on its heliocentric distance $r_{\rm
o}$. In Section \ref{Sec:HalfOneAU-compare}, we compare
observations made from radial distances of 0.5 and 1 AU. In
Section \ref{Sec:Depend-L5}, we consider two particular
observation sites that are often considered favourable in terms of
WL imaging, the L4 and L5 Lagrangian points. In Section
\ref{Sec:Depend-r}, we quantify more fully the dependence of WL
imaging on $r_{\rm o}$.

\subsection{Observing an Earth-Directed shock from the L4 and L5 Lagrangian Points} \label{Sec:Depend-L5}
The L4 and L5 Lagrangian points of the Sun-Earth system are often
considered advantageous for observing Earth-directed CMEs. There
are five Lagrangian points, all in the ecliptic, i.e., L1-L5. A
spacecraft at L1, L2, or L3 is metastable in terms of its orbital
configuration, and hence must frequently use propulsion to remain
in the same orbit. In contrast, a spacecraft at L4 or L5 is
resistant to gravitational perturbations, and is believed to be
more stable. The L4 and L5 points lie $60^\circ$ ahead of and
behind the Earth in its orbit, respectively. {\it STEREO} A
reached the L4 point in September 2009 and {\it STEREO} B reached
L5 in October 2009. The twin {\it STEREO} spacecraft were
pathfinders for future L4/L5 missions
\citep{Akioka2005,Biesecker2008,Gopalswamy2011}. A spacecraft at
either L4 or L5 can perform routine side-on imaging of
Earth-directed CMEs, and hence is of great merit for space weather
monitoring.

\begin{figure}
  \includegraphics[width=.99\textwidth]{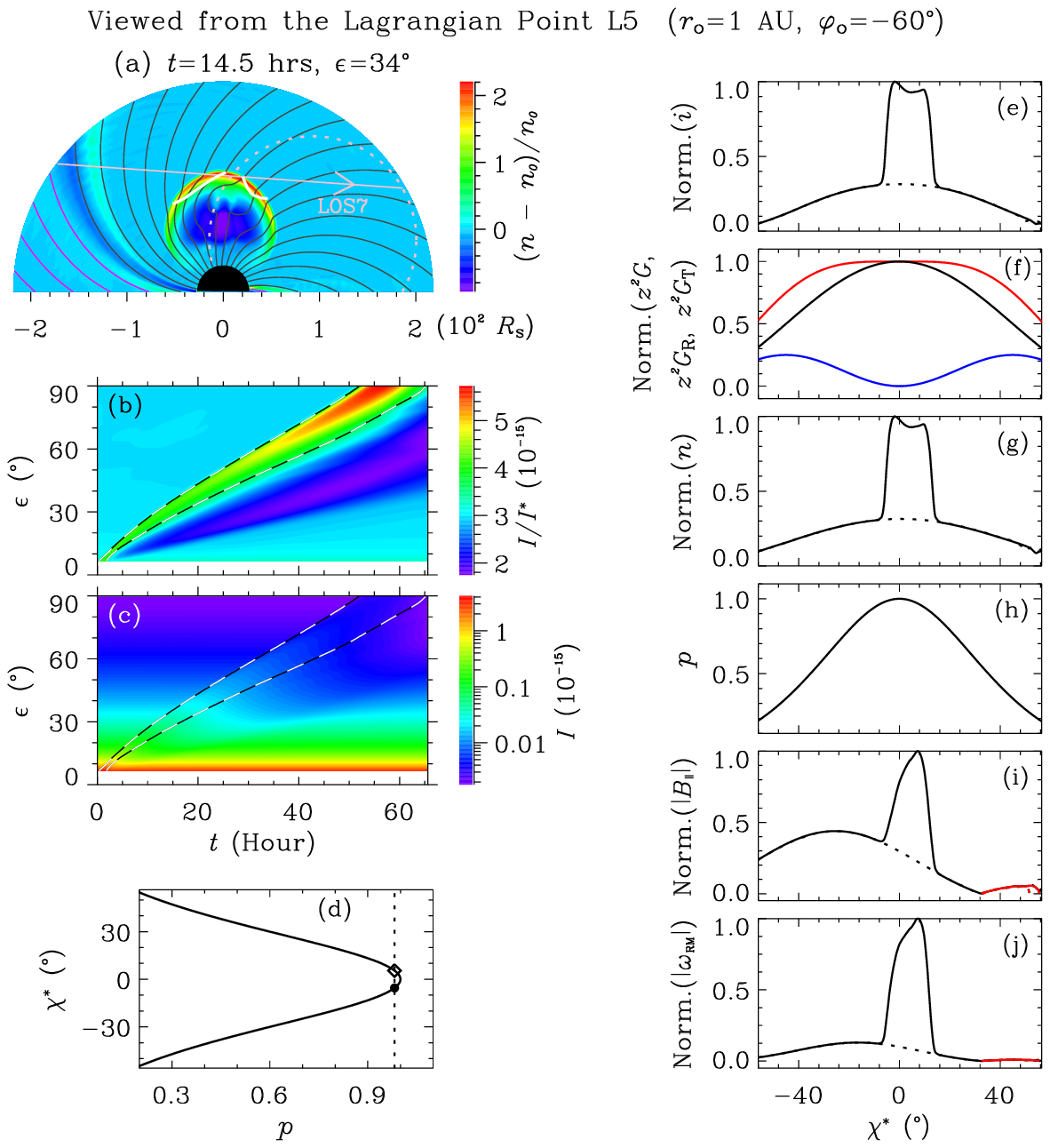}
\caption{Simulated results corresponding to the observation of an
Earth-directed shock from the L5 vantage point, i.e., along LOS7
($r_{\rm o}=1$ AU, $\varphi_{\rm o}=-60^\circ$): (panel a)
relative enhancement of electron number density $(n-n_0)/n_0$ in
the ecliptic; (panels b and c) synthesised WL time-elongation maps
of $I/I^*$ and $I$; (panel d) modified scattering angle $\chi^*$
as a function of the degree of polarisation $p$; (panels e--j) a
number of synthesised WL and FR parameters, plotted as a function
of $\chi^*$ along LOS7.} \label{60L5-contour}
\end{figure}

Figure \ref{60L5-contour}a illustrates the imaging, be it in WL or
FR, of an Earth-directed sheath from the L5 point. LOS7 intersects
the nose of the shock at an elapsed time of 14.5 hours, when the
shock nose lies on the Thomson-scattering sphere. The variation,
along LOS7, of a number of salient physical parameters is shown in
Figures \ref{60L5-contour} e--j. The interplanetary magnetic field
lines are compressed and rotated within the sheath. This rotation
results in the closer alignment of the field lines with LOS7, such
that the magnetic field component along the LOS, $|B_\parallel|$,
is greatly enhanced (Figure \ref{60L5-contour}i). The enhancements
of both $|B_\parallel|$ and electron number density $n$ within the
sheath are responsible for the resultant increases in WL radiance
$I$ and FR measurement $|\Omega_{\rm RM}|$. The degree of WL
polarisation $p$, as viewed along LOS7 that is at an elongation of
$34^\circ$, is 0.67 for the background solar wind and increases to
0.75 during the shock passage at 14.5 hours. This corresponds to a
value of the modified WL polarisation $p^*$ of 0.98, based on
Equation \ref{Equ:p*}. As was done for LOS1 in Section
\ref{Sec:HalfOneAU}, we evaluate the WL radiance along LOS7
(Figure \ref{60L5-contour}a), from which we infer the shock
position (Figure \ref{60L5-contour}d). Again, a single value of
$p^*$ corresponds to two symmetrical solutions for scattering
angle $\chi^*$, i.e., $p^*=0.29$ and $\chi^*=\pm \, 46^\circ$ for
LOS1, and $p^*=0.98$ and $\chi^*=\pm \, 5^\circ$ for LOS7. For
LOS1, only one solution for $\chi^* = - 46^\circ$ was deemed
physical; for LOS7, both solutions for $\chi^*$ are potentially
physical. The scattering sites corresponding to $\chi^*=\pm \,
5^\circ$ are very close to one another, and both agree well with
the actual position of the sheath (Figure \ref{60L5-contour}a).
The section of LOS7 bounded by $-5^\circ \le \chi^* \le 5^\circ$
lies within the sheath. Both forward scattering ($-5^\circ \le
\chi^* < 0^\circ$) and backward scattering ($0^\circ < \chi^* \le
5^\circ$) will contribute to the radiance $I$ observed along this
LOS. The propagating sheath can be tracked continuously and easily
in WL from the L5 vantage point, such that it leaves an obvious
signature in the J-map of synthesised radiance (Figures
\ref{60L5-contour}b and \ref{60L5-contour}c). This confirms
previous assertions that the L4 and L5 points are very favourable
in terms of space weather monitoring.

\subsection{Dependence of White-Light Radiance on Heliocentric Distance} \label{Sec:Depend-r}
\begin{figure}
 \includegraphics[width=.5\textwidth]{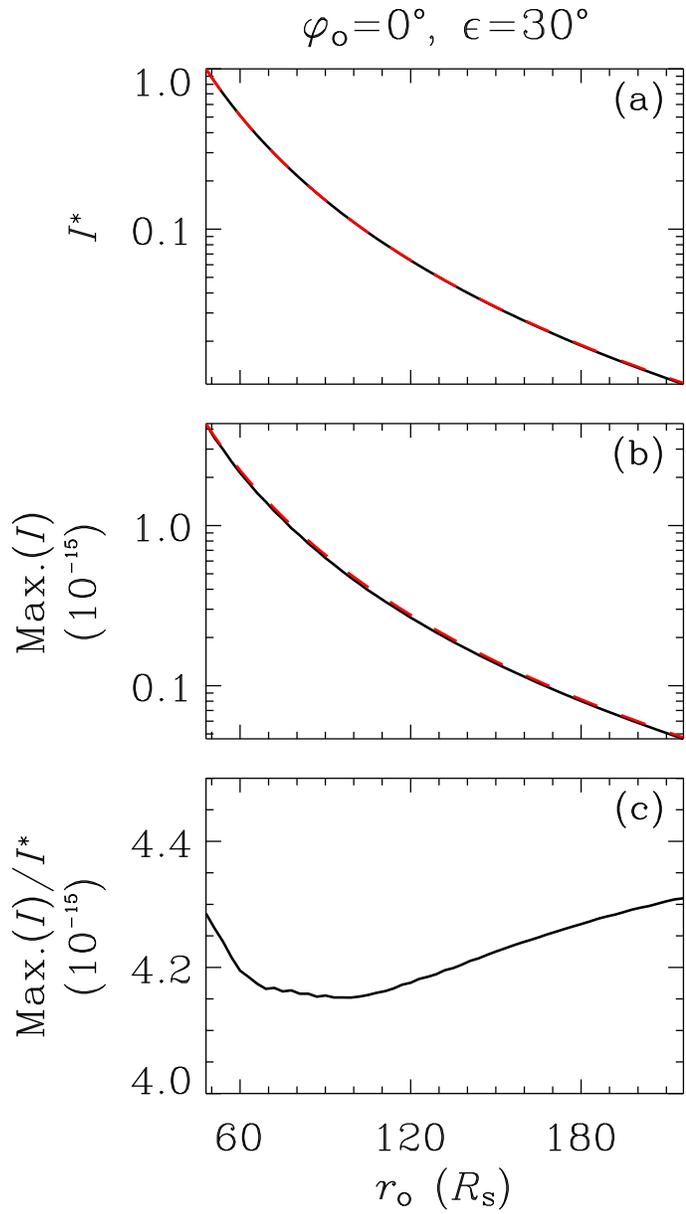}
\caption{Simulated WL radiance, as a function of $r_{\rm o}$,
viewed along an elongation of $30^\circ$ by an observer on the
Sun-Earth line ($\varphi_{\rm o}=0^\circ$, $50 \, R_{\rm S} \le r
\le 215 \, R_{\rm S}$). Again, ``Max" refers to the strongest
signal associated with the sheath passage. The dashed red lines in
the two upper panels show, for comparison, an $r^{-3}$ variation.
Both $I^*$ and $\mbox{Max.}(I)$ are well described by such a
variation.} \label{90Line-r}
\end{figure}

The background intensity at a fixed elongation in a WL sky map is
greater for an observer closer to the Sun. For a heliospheric
imager at any distance from the Sun, \citet{Jackson2010} proposed
the following Thomson-scattering principles: (1) The WL radiance
$I$ at a given solar elongation falls off with the heliocentric
distance $r$ according to $r^{-3}$; (2) Such a dependence of $I
\propto r^{-3}$ is valid for almost any viewing elongation, and
for any radial distance from 0.1 AU out to 1 AU and beyond. The WL
radiance $I$ depends on the heliospheric distribution of electron
number density $n$. In interplanetary space, the background solar
wind speed is nearly constant, and the background electron number
density $n_0$ varies approximately with $r^{-2}$. However, the
equilibrium defined by $n_0 \propto r^{-2}$ is disturbed by the
presence of interplanetary transients, such as CMEs and CIRs. A
travelling shock can sweep up, and hence compress significantly,
the background solar wind plasma. Figures \ref{10Contour} and
\ref{60L5-contour}a show a density enhancement of
$\frac{n-n_0}{n_0} \approx 2.2$ within the sheath. The associated
compression ratio $\frac{n}{n_0} \approx 3.2$ indicates that the
shock is very strong. However, when viewed along elongations less
than $60^\circ$, the strongest signatures of shock passage
(characterised by Max.($I$)) vary very closely with $r^{-3}$
(Figures \ref{90Line-r}b and \ref{90Line-r}c). The relationship of
Max.($I$) $\propto  r^{-3}$ is slightly violated at large
elongations $\varepsilon > 60^\circ$. Figure \ref{50Compare}c
reveals that the ratio between Max.($I_{\rm 0.5 AU}$) and
Max.($I_{\rm 1 AU}$) is close to 8 for $\varepsilon \le 60^\circ$,
increasing thereafter to 10.8 at $\varepsilon = 180^\circ$. The
premise that the WL radiance decreases with the third power of
Sun-observer distance generally holds true for both the background
solar wind and propagating CMEs.

%------------------------------------------------------------------------------------------------------------------
\section{Discussion}\label{Sec:discussion}
The detectability in WL of a particular electron density feature
is determined by its signal above the noise background. In {\it
STEREO}/HI-1 images, the dominant WL signal is zodiacal light due
to scattering of sunlight from the F-corona, which is centred
around the ecliptic. In the {\it STEREO}/HI-2 FOV, the noise floor
is primarily determined by photon noise and the background
star-field \citep{DeForest2011}. Away from the ecliptic, the
background WL noise has a sharp radial gradient in coronal images,
and is nearly constant in heliospheric images. The signal-to-noise
ratio for heliospheric electron density features is discussed by
\citet{Howard2013}. We will address the detection of CMEs in the
presence of background noise in future forward-modelling work.

If both a transient CME and background (Heliospheric Current Sheet
(HCS) - Heliospheric Plasma Sheet (HPS)) plasma structures are
present along the same LOS, both will contribute to the total
LOS-integrated radiance. In this case, the interpretation of the
data would clearly be more problematic. Moreover, if the LOS were
to penetrate a HCS, the magnetic field vector would, at that
point, rotate through 180$^\circ$. Due to the mutual cancellation
of $B_\parallel$ across the HCS, there may be no net FR signature
according to Equations \ref{Equ:FR1}-\ref{Equ:FR3}. Hence, even
such a significant interplanetary structure may be associated with
only a weak FR measurement. Conversely, the relatively dense
plasma within a HPS can significantly contribute to WL radiance.
Thus the potential effects of the presence of HCS-HPS structures
need to be borne in mind in the remote imaging of CMEs. In the
current work, however, we find that such effects are negligible.
In our numerical simulation, there are two HCS-HPS structures,
which are initially rooted at longitudes of $\varphi=\pm \,
90^\circ$ at the inner boundary of our numerical simulation. The
simulated shock emerges at a longitude of $\varphi=0^\circ$. The
large longitudinal difference between the HCS-HPS and the shock
means that the remote-sensing signatures are principally
contributed by the sheath. Thus, in our forward-modelling work,
the signal enhancements of synthesised imaging in WL and FR are
unambiguously the result of the propagating sheath.

In general, the more complex the interplanetary dynamics, the more
complex the resultant remote-sensing observations will be. For
instance, a CME can interact with other CMEs and/or background
solar wind structures such as CIRs, HCSs, and HPSs; mutual
interaction between CMEs is, however, generally more perturbing
than interactions between CMEs and such background structures.
Interactions can result in the background solar wind structures
becoming warped or distorted \citep[e.g.,][]{Odstrcil1996,Hu2001},
and CMEs being accelerated/decelerated
\citep[e.g.,][]{Lugaz2005,Xiong2007,Shen2012b}, deflected
\citep[e.g.,][]{Xiong2006b,Xiong2009,Lugaz2012}, distorted
\citep[e.g.,][]{Xiong2006b,Xiong2009}, or entrained
\citep[e.g.,][]{Rouillard2009a}. In particular, during such
interactions, the behavior of a sheath can become much more
complex: the shock aphelion can be deflected, spatial asymmetries
can develop along the shock front, and the shock front can
potentially merge completely with other shock fronts. At an
interaction site, both the plasma density and magnetic field would
be compressed; this would lead to enhanced signatures in both WL
and FR observations. For example, the interaction between two CMEs
was manifest as a very bright arc in WL images
\citep[e.g.,][]{Harrison2012,Liu2012,Temmer2012}. Different types
of interaction would likely result in different WL radiance
signatures; in fact, through a single interaction, the
corresponding radiance pattern would evolve. The interpretation of
such complex WL radiance patterns would be prone to large
uncertainties, but can be rigorously constrained if interplanetary
imaging was performed from multiple vantage points and
complemented by numerical modelling. For stereoscopic WL imaging,
ray-paths from one observer intersect those from the other
observer. Thus the 3D distribution of electrons in the inner
heliosphere can be reconstructed using a time-dependent tomography
algorithm \citep{Jackson2006,Bisi2008,Webb2013}. With the aid of
numerical modelling, coordinated imaging in WL and FR would enable
the properties and evolution of complex interplanetary dynamics to
be diagnosed.

\section{Concluding Remarks} \label{Sec:conclusion}
In this paper, we have investigated the WL and FR signatures of an
interplanetary shock based on an approach of forward MHD
modelling. The WL Thomson-scattering geometry is increasingly more
significant at larger elongations. The degree of WL polarisation
can be used to estimate the 3D location of the main scattering
region, while FR measurement can be used to infer, to some extent,
the magnetic configurations of CMEs. This work presented here
demonstrates, as a proof-of-concept, that the availability of
coordinated observations in polarised WL and FR measurement would
enable the local LOS magnetic component to be estimated. Although
the current generation of heliospheric WL imagers, such as the
{\it STEREO}/HI instruments, do not have polarisers, there are
advances underway in terms of FR imaging using Low-frequency radio
arrays. Coordinated imaging in WL and FR would enable the inner
heliosphere to be mapped in fine detail; the location, mass, and
magnetic field of CMEs can be diagnosed on the basis of such
combined observations. Forward modelling is crucial in
establishing the causal link between interplanetary dynamics and
observable signatures, and can provide valuable guidance for
future coordinated WL and FR imaging.

Although not the methodology of the current work, numerical MHD
models of the inner heliosphere can also be directly driven by
photospheric observations
\citep[e.g.,][]{Hayashi2005,Wu2006,Feng2012b}. A comparison of
synthesised and observed WL and FR sky maps, the former based on
the use of such data-driven models, would prove highly beneficial
in validating the forward modelling and interpreting the
observations. Such an integration of observation data analysis and
numerical forward modelling will be explored as the continuation
of the preliminary modelling work presented in this paper.

\acknowledgments This work is jointly supported by the National
Basic Research Program (973 program) under grant 2012CB825601, the
Chinese Academy of Sciences (KZZD-EW-01-4), the National Natural
Science Foundation of China (41231068, 41031066, 41204129), the
Strategic Priority Research Program on Space Science from the
Chinese Academy of Sciences (XDA04060801), the Specialized
Research Fund for State Key Laboratories of China, the Chinese
Public Science and Technology Research Funds Projects of Ocean
(201005017), open research foundation of Science and Technology on
Aerospace Flight Dynamics Laboratory of China (AFDL2012002),
research fund for recipient of excellent award of the Chinese
Academy of Sciences President's scholarship (startup fund). Ming
Xiong is also partially supported by an institutional project of
``Key Fostering Direction in Pulsar Science and Application" from
the Center of Space Science and Applied Research, China. Ming
Xiong sincerely thanks Drs. Bo Li, Ding Chen, Craig DeForest,
James Tappin, and Tim Howard for their beneficial discussions and
thoughtful suggestions. We sincerely thank the anonymous referee
for his/her constructive suggestions.

\begin{table}
\begin{center}
\rotatebox{90}{
\begin{minipage}{240mm}
\begin{center}
\caption{Maximum values of the parameters $i$, $z^2 G$, $z^2
G_{\rm R}$, $z^2 G_{\rm T}$, $n$, $|B_\parallel|$, and
$|\omega_{\rm RM}|$ along LOS1, LOS2, LOS3, and LOS7, used for
normalisation in Figures \ref{20LOS-a} and \ref{60L5-contour}e-j.
Each LOS is designated a time $t$, a radius $r_{\rm o}$, a
longitude $\varphi_{\rm o}$, and an elongation $\varepsilon$ in
columns 2-5; LOSs 1-6 are overlaid on Figure \ref{10Contour} and
LOS7 is overlaid on Figure \ref{60L5-contour}.}\label{Tab:LOS}
\begin{tabular}{cccccccccc} \hline %
LOS & Time   & Radii       & Longitude         & Elongation    & WL radiance        & WL Thomson-scattering                     & Electron    & Parallel magnetic & FR measurement \\ %
    & $t$    & $r_{\rm o}$ & $\varphi_{\rm o}$ & $\varepsilon$ & $i$                & geometry factors                           & number      & field             & \\ %
    & (hour) & (AU)        & ($^\circ$)        & ($^\circ$)    & ($\times 10^{-27}$)& $z^2 G$, $z^2 G_{\rm R}$, $z^2 G_{\rm T}$  & density $n$ & $|B_\parallel|$   & $|\omega_{\rm RM}|$\\ %
    &        &             &                   &               &                    & ($\times 10^{-29}$)                        & (cm$^{-3}$) & (nT)              & ($\times 10^{-12}$ rad m$^{-3}$) \\ \hline %
LOS1 & 5.5  & 0.5 & 0   & 20  & 36.3  & 9.33  & 392   & 207  & 14.4  \\ \hline %
LOS2 & 5.5  & 1   & 0   & 7   & 344   & 18.2  & 1893  & 223  & 18.2  \\ \hline %
LOS3 & 14   & 1   & 0   & 20  & 1.93  & 2.31  & 87    & 68.9 & 1.51  \\ \hline %
LOS7 & 14.5 & 1   & -60 & 34  & 0.52  & 0.86  & 60.3  & 28.5 & 0.42  \\ \hline %
\end{tabular}
\end{center}
\end{minipage}
}
\end{center}
\end{table}
\end{document}